\documentclass[preprint]{aastex}

\newcommand{\dsct}{$\delta$\,Sct}
\newcommand{\ds}{$\delta$\,Sct}
\newcommand{\gdor}{$\gamma$\,Dor}
\newcommand{\Msun}{$M_{\odot}$}

\newcommand{\Rsun}{$R_{\odot}$}

\newcommand{\Hg}{H$_{\gamma}$}
\newcommand{\Tf}{$T_{\mathrm{eff}}$}
\newcommand{\lgg}{log g}
\newcommand{\hda}{HD\,114839}
\newcommand{\bda}{BD\,+18\,4914}
\newcommand{\vsini}{\ensuremath{{\upsilon}\sin i}}
\newcommand{\kms}{$\mathrm{km\,s}^{-1}$}
\newcommand{\cms}{$\mathrm{cm\,s}^{-2}$}
\newcommand{\cd}{\,$\rm{d}^{-1}$}
\newcommand{\gdhybrid}{\dsct\ - \gdor\ hybrid}
\newcommand{\vmic}{$\upsilon_{\mathrm{mic}}$}
\newcommand{\vrad}{${\upsilon}_{\mathrm{r}}$}

\newcommand{\atlas}{{\sc ATLAS9}}
\newcommand{\synth}{{\sc SYNTH3}}
\newcommand{\vald}{{\sc VALD}}
\newcommand{\errvr}{$\sigma_{{\upsilon}_{\mathrm{r}}}$}
\newcommand{\errTeff}{$\sigma_{T_{\mathrm{eff}}}$}
\newcommand{\errlogg}{$\sigma_{\ensuremath{\log g}}$}

\shorttitle{Abundance analysis of two hybrid pulsating stars: \hda\ and \bda}
\shortauthors{Hareter et al.}

\begin{document}

\title{Looking for a connection between the Am phenomenon and hybrid \ds\,-\gdor\ pulsation: a determination of the fundamental parameters and abundances of HD\,114839 and BD\,+18\,4914\thanks{Based on observations made at the Observatoirede Haute-Provence and with HARPS at the 3.6m ESO telescope under the Large Programme LP185.D-0056.} }

\author{M. Hareter} 
\affil{University of Vienna, Department of Astronomy, T\"urkenschanzstrasse 17, A-1180 Wien, Austria}
\email{hareter@astro.univie.ac.at}

\author{L. Fossati} 
\affil{Department of Physics and Astronomy, Open University, Walton Hall, Milton Keynes MK6 7AA, United Kingdom}

\author{W. Weiss} 
\affil{University of Vienna, Department of Astronomy, T\"urkenschanzstrasse 17, A-1180 Wien, Austria}

\author{J. C. Su\'arez} 
\affil{Instituto de Astrof\'isica de Andaluc\'ia (CSIC), Glorieta de la Astronom\'{\i}a S/N, 18008, Granada, Spain}

\author{K. Uytterhoeven} 
\affil{Kiepenheuer-Institut f\"ur Sonnenphysik, Sch\"oneckstr. 6, D-79104 Freiburg, Germany; Instituto de Astrof\'isica de Canarias, 38200 La Laguna, Tenerife, Spain; Departamento de Astrof\'isica, Universidad de La Laguna, 38205  La Laguna, Tenerife, Spain}
\and
\author{M. Rainer} 
\author{E. Poretti} 
\affil {INAF-Osservatorio Astronomico di Brera, Via E. Bianchi 46, 23807, Merate (LC), Italy}

\begin{abstract}
\gdhybrid s pulsate simultaneously in p- and g-modes, which carry information on the structure of 
the envelope as well as to the core. Hence they are key objects for investigating 
A and F type stars with asteroseismic techniques. 
An important requirement for seismic modeling are small errors in temperature, gravity and chemical composition.
Furthermore, we want to investigate the existence of an abundance indicator typical for hybrids, 
something that is well established for the roAp stars. Previous to the present investigation, the abundance 
pattern of only one hybrid and another hybrid candidate has been published.
We obtained high-resolution spectra of \hda\ and \bda\ using the SOPHIE 
spectrograph of the Observatoire de Haute-Provence and the HARPS spectrograph
at ESO La\,Silla. For each star we determined fundamental parameters and 
photospheric abundances of 16 chemical elements by comparing synthetic 
spectra with the observations. We compare our results to that of seven \ds\ and
 nine \gdor\ stars. 
For the evolved \bda\ we found an abundance pattern typical for an Am star, 
but could not confirm this peculiarity for the less evolved star \hda, which is classified 
in the literature as uncertain Am star. Our result supports the concept of evolved 
Am stars being unstable. With our investigation we nearly doubled the number of spectroscopically 
analyzed \gdhybrid\ stars, but did not yet succeed in identifying a spectroscopic 
signature for this group of pulsating stars.
A statistically significant spectroscopic investigation of \ds- \gdor\ hybrid stars 
still is missing, but would be rewarding considering the asteroseismological potential 
of this group.
\end{abstract}

 
\section{Introduction}
\ds\ stars are main sequence or evolved stars (up to luminosity class III), 
which pulsate in low-order radial and non-radial p-modes with typical pulsation 
frequencies of 5 to 50\,\cd. They occupy the lower Cepheid instability strip 
extended towards the main sequence \citep[see][for a detailed review]{Breger00}. 
The class of \gdor\ stars was defined in 1999 by \citet{Kaye99} and, in contrast 
to \ds\ stars, they pulsate in high-order g-modes with typical frequencies of 0.3 
to 3\,\cd. The instability strip of \gdor\ stars is rather narrow and partly overlaps 
the larger \ds\ instability strip. Given the overlap between the \ds\ and \gdor\ 
instability strips, a search for hybrid pulsators began in the last few years, 
which was driven by the fact that p-modes are sensitive to the envelope, whereas g-modes 
are sensitive to the core. Hence hybrids are key objects for testing stellar models 
from core to the photosphere with asteroseismic techniques. 

The space missions MOST \citep[Microvariability and Oscillation of STars,][]{Walker03},
CoRoT \citep[COnvection, ROtation and planetary Transits,][]{esa3}, and {\it Kepler} \citep{Borucki2010} 
increased the number of known hybrid stars significantly, because observations of low frequency
pulsation with low amplitudes are easier from space \citep[see e.g.][]{King06, Rowe06, Hareter10, Grig10, Uytterhoeven11}.

Am stars are chemical peculiar early F- to A-type stars, characterized by
an underabundace of C, N, O, Ca and Sc, as well as by a general
overabundance of Fe-peak and Rare Earth Elements. Am stars do not host
large-scale magnetic fields \citep{auriere10} and are often in binary
systems (see e.g. recent review by \citet{Iliev2008} and references therein). 
Compared to the vast majority of the chemically normal stars of the same temperature, Am stars are slow rotators, allowing diffusion processes
to produce the chemical peculiarities \citep{Michaud70}.

The aim of this work is to investigate a potential link between chemical peculiarities such as metallic line A/F stars 
to the hybrid pulsation. Furthermore, we intend to provide accurate fundamental parameters of our two stars for further 
modeling of this type of pulsator.

With HD\,8801 \citet{Henryfekel05} discovered the first \ds-\gdor\ hybrid. Based on the Ca K 
line intensity relative to hydrogen and metallic lines they classified HD\,8801 as an Am star. 
Hence it was obvious to check if the two \gdhybrid s discovered by \citet[\hda]{Rowe06} and by \citet[\bda]{King06} 
with MOST photometry share this chemical peculiarity.

\hda\ (RA: 13\fh12\fm47\fs43, and DEC: +26\fdg52\arcmin52\farcs19, epoch 2000; $V$=8.46\,mag) 
was classified by \citet{Hill76} and \citet{Clausen79} as an uncertain Am 
star. \citet{Pribulla09} analyzed two spectra obtained at the David Dunlap 
Observatory (DDO) in the spectral region of the \ion{Mg}{1} triplet at $\lambda\lambda\sim$5167, 5173, 
and 5184\,\AA, from which they deduced that \hda\ is a metallic line star of spectral type F4/5. 
They measured a \vsini\ of 70\,\kms, low enough to allow \hda\ to be an Am star \citep{charbonneau}.
\bda\  (RA: 22\fh02\fm37\fs70 and DEC: +18\fdg54\arcmin02\farcs62, epoch 2000) is a fainter star 
($V$=10.6\,mag), for which the MOST photometry \citep{Rowe06} showed clearly two distinct groups of frequencies. 
This star was also included in the survey by \citet{Pribulla09}, who derived a spectral type of F5/6 
and a \vsini\ of 40\,\kms. 

As a side comment we want to mention HD\,49434, for which a ground-based photometric and spectroscopic campaign was 
organized and published by \citet{Uytterhoeven08}. However, the hybrid nature of this star has not yet been established beyond doubt, 
because this star might rotate fast and a spread of frequencies over a larger range may mimic hybrid characteristics. 
The abundance analysis for this star did not show significant peculiarities.


\section{Observations and data reduction}

Four spectra of \hda\ were obtained between 2008, April 13 and May 2, 
while two spectra of \bda\ were obtained in 2008, August 7 and 8 
with the SOPHIE spectrograph. 

SOPHIE is a cross-dispersed \'echelle spectrograph mounted at the 1.93-m 
telescope of the OHP.  The spectra cover the wavelength range of
3872--6943\,\AA\ without gaps and we used it in the medium spectral resolution 
($R\sim$40\,000) mode.  

The orders of the individual SOPHIE spectra were averaged,  the blaze function removed and the individual orders merged to a single 
spectrum. The lowest quality spectrum of \hda\ was ignored in this process. 
The removal of the blaze function containing hydrogen lines (\hda\ and \bda\ in our case) and the subsequent continuum normalization is a challenge for spectra 
obtained with \'echelle spectrographs, because for A to F stars these lines extend over adjacent orders. We corrected for the blaze 
function of orders containing the hydrogen lines by using the artificial flat-fielding 
technique described, e.g., by \citet{barklem02}. This approach assumes that the 
blaze shape of the different \'echelle orders change smoothly with the order 
number and one can reconstruct missing blaze shapes by fitting polynomials to 
continuum points identified in several adjacent orders blue- and redwards of the 
hydrogen line containing order. A 2-D polynomial surface is fitted to these empirical blaze functions and the blaze in the orders containing the hydrogen 
lines is determined by 
interpolation. The accuracy of this technique is confirmed by 
an excellent agreement of the normalized hydrogen lines obtained with SOPHIE and 
HARPS spectrographs (see Fig.\,\ref{fig-hda-hbetafit}) having a clearly different blaze characteristic and different position of the hydrogen lines within the orders.  

The \vsini\ of the two stars ($\sim$70 and 40\,\kms) makes the continuum 
normalization a critical step of the reduction procedure. We 
normalized the combined, blaze removed orders without the use of any automatic continuum fitting procedure and by comparing with suited synthetic spectra. 
For wavelengths shorter than the \Hg\ line it 
was not possible to determine the correct level of the continuum, as there 
were not enough continuum windows in the spectrum at these wavelengths 
due to the strong line blending. Therefore we did not include this part of 
the spectrum in our analysis.
Finally, we obtained a signal-to-noise ratio (S/N) per pixel of 121 at $\sim$5800\,\AA\ for \hda\ and of 56 for \bda. 

Two spectra of \hda\ were obtained with HARPS two years later on June 21, and 22, 2010, and one of \bda\ in the same year on July 2, in the framework of a 
programme complementary to the extensive monitoring of the hybrid \ds\ - \gdor\ variables observed with the CoRoT satellite. We processed the HARPS data 
in a manner similar to the SOPHIE spectra and achieved a S/N per wavelength pixel for \hda\ of 200 at $\sim$5800\,\AA\  
and for \bda\ of 92. The HARPS spectra were obtained in the 
EGGS mode, which delivers a resolution of 80\,000.

Table~\ref{tab-spacings} presents the observing log with our measured stellar 
radial velocities (\vrad). The star \bda\ might be the primary star 
in a binary system, given the difference in \vrad\ obtained between the 
SOPHIE and HARPS spectra in 2008 and 2010. However, from ground based observations 
\citep{Mathias04} it is known that \gdor\ stars 
may show significant RV variations due to pulsation. 
None of our spectra shows spectral lines from a secondary, hence the assumption of a 
single star or a single lined binary is justified.

\begin{table}

\caption{Observing log of the SOPHIE and HARPS observations with measured radial velocity values and the errors.}

\begin{center}
\begin{tabular}{lccccc}

\tableline\tableline
Star  & BJD $-$ & S/N & Exp. time & \vrad & \errvr \\
Name  & 2450000 &     &  s        & \kms & \kms \\
\tableline
\hspace{3mm} \textit{SOPHIE:}\\
\hda\  & 4570.48314 &  72 & 1500 & -7.5 & 3.6 \\
\hda\  & 4571.46797 & 114 & 1500 & -7.4 & 2.7 \\
\hda\  & 4573.62490 &  44 & 1500 & -7.4 & 3.2 \\
\hda\  & 4589.51983 & 111 & 1500 & -8.0 & 2.6 \\
\bda\  & 4685.53476 &  50 & 2400 & -42.3 & 1.4 \\
\bda\  & 4686.45132 &  51 & 2400 & -42.8 & 1.6 \\
\tableline
\hspace{3mm}\textit{HARPS:} \\
\hda\  & 5368.55188 & 198 & 2400 & -8.2 & 2.5 \\
\hda\  & 5369.53145 & 127 & 2400 & -7.0 & 2.2 \\
\bda\  & 5379.92632 &  92 & 2500 & -47.0 & 0.9 \\
\tableline

\end{tabular}
\end{center}
\smallskip 

The S/N per pixel are calculated at $\sim$5800\,\AA. Exposure times are given 
in seconds. BJD indicates the Barycentric Julian Date at the middle of 
the exposure.

\label{tab-spacings}
\end{table}

\section{Parameter Determination and Abundance Analysis}\label{s:param}

To compute model atmospheres we employed a modified version of \atlas\  
\citep{Kurucz93b}, which uses a model of convection based on \citet{cm1,cm2}. 
More details can be found in \citet{heiter}. The initial atmospheric parameters for \hda\ were derived from Str\"omgren 
photometry by \citet{HauckM} employing the calibration by \citet{Balona94}.
We obtained \Tf\ = 7400\,K and \lgg\ = 4.20. For \bda\ no narrow band photometry
was available. Based on a comparison of the spectra we decided to use the same 
parameters as those used for \hda.

\begin{figure}
    \epsscale{0.70}
    \plotone{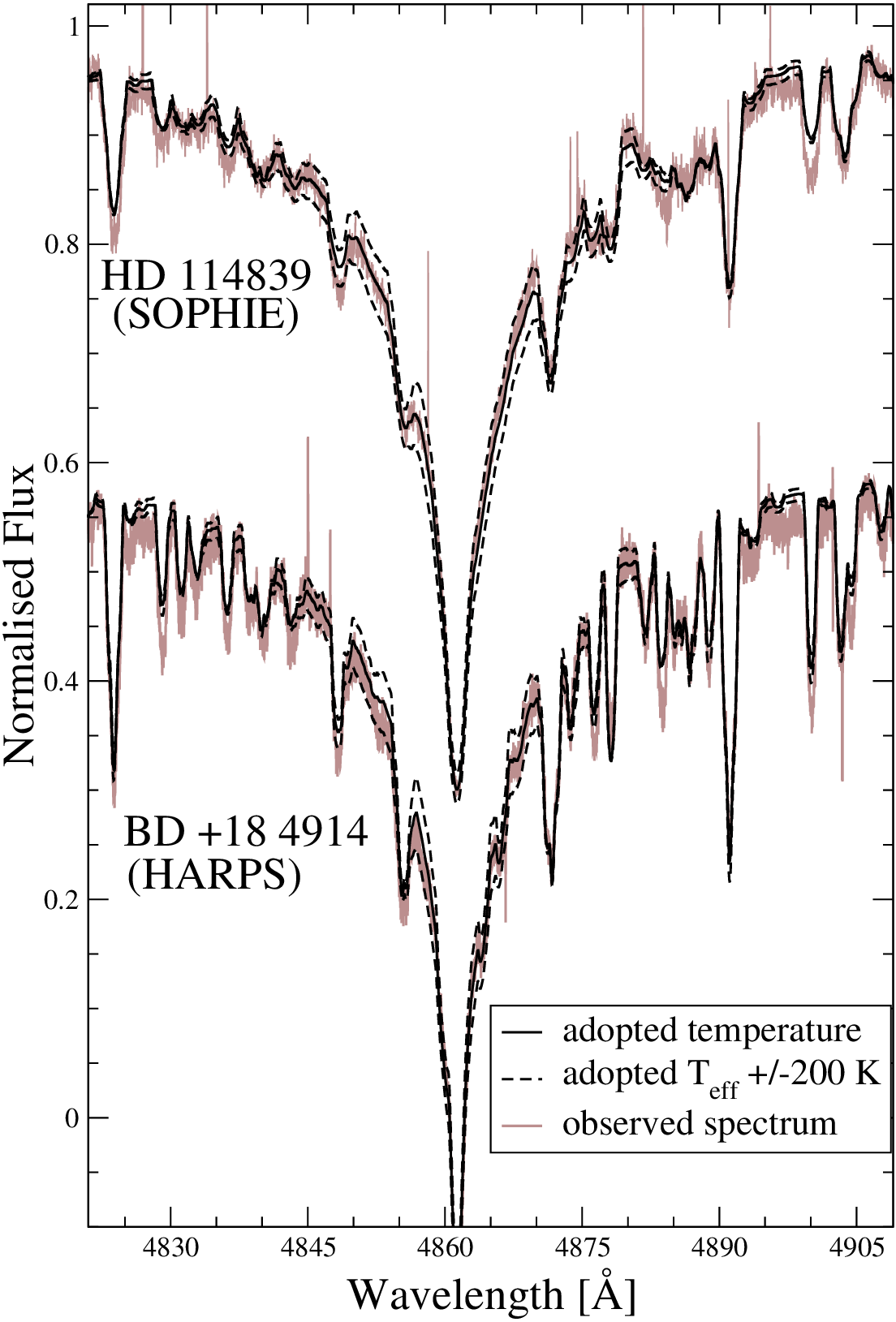}
 \caption{Comparison of the H$\beta$ line profiles (wide gray solid lines) 
 to synthetic profiles calculated with the final adopted \Tf\ (solid lines), 
 and uncertainty (dashed lines).}  
 \label{fig-hda-hbetafit}
\end{figure}

In the temperature range of \hda\ and \bda, the hydrogen line wings 
are sensitive to \Tf, making them very good temperature 
indicators. We fitted synthetic hydrogen line profiles, calculated with 
\synth\ \citep{Kochukhov07}, to the observations. Figure~\ref{fig-hda-hbetafit} shows the comparison of the observed 
(SOPHIE for \hda, top, and HARPS for \bda, bottom) H$\beta$ line profiles for 
the two stars with synthetic profiles calculated with the stellar 
\Tf\ value of 7100\,K for \hda\ and 6900\,K for \bda. We also 
show synthetic line profiles calculated with a \Tf\ differing by 
1$\sigma$ ($\pm$ 200\,K).

The \ion{Fe}{1} and \ion{Fe}{2} excitation equilibrium confirms the \Tf\ obtained from the 
hydrogen line fitting. The atomic line parameters were obtained from the \vald\ 
database \citep{vald1,vald2,vald3}. By requiring ionization equilibrium for 
\ion{Fe}{1} and \ion{Fe}{2} we derived \lgg\ values of 4.0 +/- 0.2 and 3.8 +/- 0.2 
for \hda\ and \bda, respectively. 

\citet{vanLeeuwen07} reported for \hda\ a parallax 
of 6.24$\pm$0.85\,mas, which puts the star at a distance between 140 and 185\,pc.
The bolometric correction of 0.033 \citep[see][]{Flower96}. was neglected, because 
the uncertainty of the parallax plays a much larger role. 
From this distance and making use of isochrones by \citet{marigo} (see Sect.~\ref{hrd}), we derived a 
stellar gravity of 4.00$^{+0.14}_{-0.12}$, in agreement with what we obtained
by imposing the ionization equilibrium.

The median radial velocities and \vsini\ values were derived from fitting synthetic 
spectral lines to observed profiles. Our measured \vrad\ values are listed in Table~\ref{tab-spacings} 
and we obtained \vsini\ values of 68$\pm$2\,\kms\ and 38$\pm$2\,\kms, respectively, for \hda\ and \bda . 

\begin{figure}
    \epsscale{0.70}
    \plotone{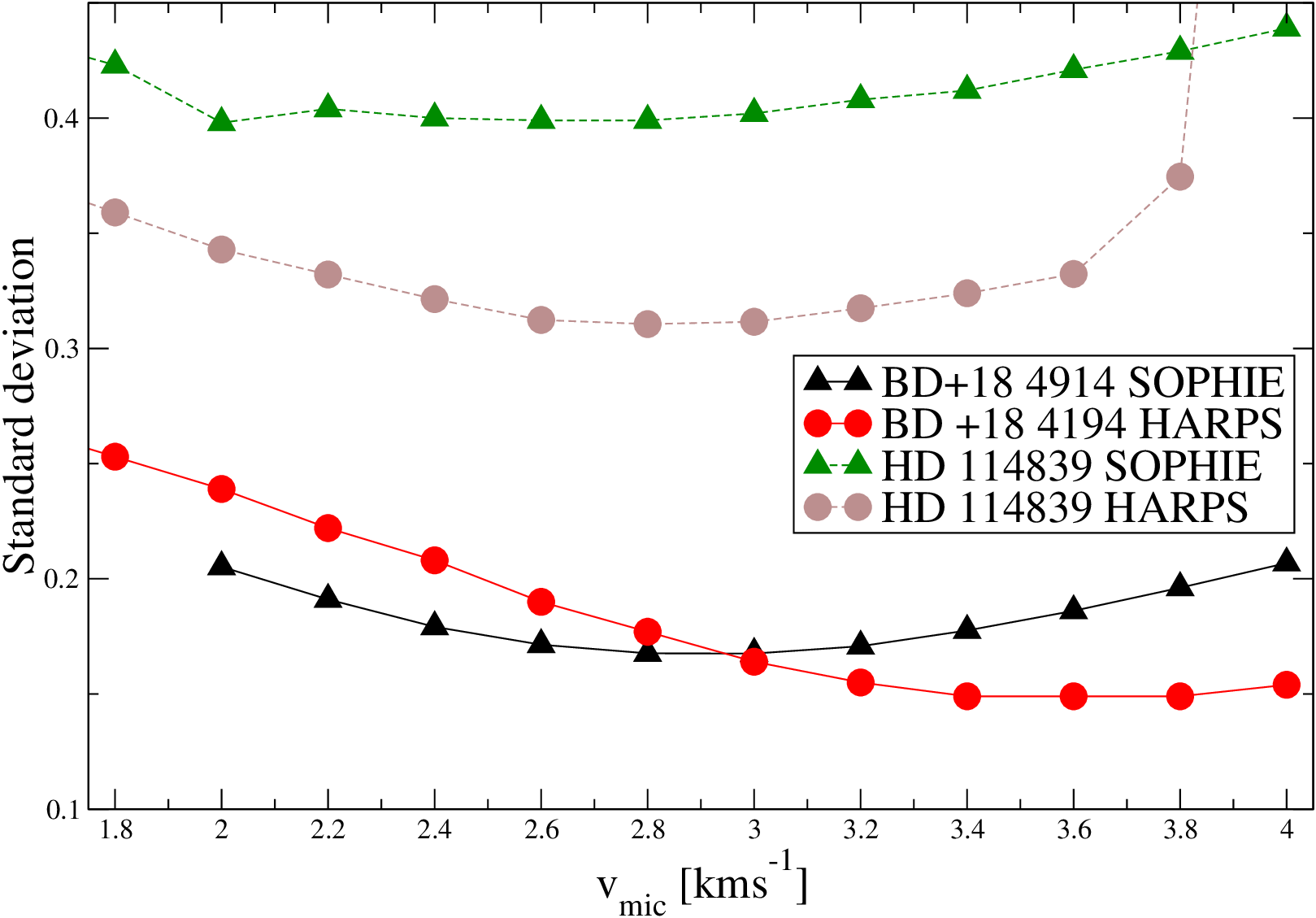}
 \caption{Variation of the standard deviation from the mean Fe abundance of
 both stars as function of \vmic, as obtained from the SOPHIE (triangles) 
 and HARPS (circles) spectrum. Dashed lines indicate measurements for \hda\ (offset by 0.2 for better visibility), while full lines
 connect measurements for \bda.}
 \label{fig-hda-vmic}
\end{figure}

Due to the rather high \vsini\ of both stars, it was not possible to reliably measure equivalent 
widths for the metallic lines. Therefore we derived the line abundances by line core 
fitting in the same way as described by \citet{Fossati08}. The microturbulence 
velocity was determined using only Fe lines, because the high \vsini\ and rather 
low S/N did not allow us to measure enough lines of other elements. We derived the line 
abundances for the selected Fe lines at different \vmic\ values and plotted the standard 
deviation from the mean abundance as a function of \vmic\ (Figure~\ref{fig-hda-vmic}). 
The minimum of this curve gives the \vmic\ value, which fits the observation best. From the 
SOPHIE spectra of \hda\ and \bda\ we derived \vmic\ of 2.6 and 2.9\,\kms, respectively, with 
an uncertainty of 0.7\,\kms, while from the HARPS spectra we derived 
\vmic\ of 2.8 and 3.6\,\kms, respectively, with an uncertainty of 0.7\,\kms.
The values agree within the quoted errors, however, we remark that the HARPS spectra have a better S/N ratio.

The abundances we obtained for the two stars are listed in Tables~\ref{tab-hda-abundances} 
and \ref{tab-bda-abundances}. We also checked for rare earth element overabundances, 
but none was found. Due to the presence of artifacts in the SOPHIE spectra at wavelengths 
close to the \ion{Li}{1} line at $\lambda$6707\,\AA, we were not able to measure the Li 
abundance from these spectra. The analysis of the Li line was done with the HARPS spectra.

\begin{table}
\caption{LTE element abundances, in $\log (N/N_{\rm tot})$, for \hda\ determined 
from SOPHIE (second column) and HARPS (third column) spectra. 
The estimated internal errors in units of 0.01\,dex and the number of lines 
measured for each element are given in parenthesis. For comparison, the last 
column shows the solar abundances by \citet{asplundetal2005}. Abundances 
obtained from just one line have no error (-;1).}
\begin{center}
\begin{tabular}{lccc}
\tableline\tableline
        & \multicolumn{3}{c}{$\log (N/N_{\rm tot})$} \\ 
Element & SOPHIE & HARPS & Sun \\
\tableline
Li & -(-;0)       & -8.91(-;1)   & -10.99 \\
C  & -3.36(-;1)   & -3.36(-;1)   & -3.65  \\ 
Na & -5.82(16;2)  & -5.68(11;2)  & -5.87  \\
Mg & -4.44(27;9)  & -4.55(12;5)  & -4.51  \\
Si & -4.22(21;6)  & -4.37(23;8)  & -4.53  \\
S  & -4.60(-;1)   & -4.75(12;4)  & -4.90  \\
Ca & -5.81(22;19) & -5.81(12;12) & -5.73  \\
Sc & -9.11(27;6)  & -9.11(05;6)  & -8.99  \\
Ti & -7.16(31;10) & -7.10(34;9)  & -7.14  \\
Cr & -6.37(21;15) & -6.37(28;10) & -6.40  \\
Mn & -6.63(16;2)  & -6.75(01;2)  & -6.65  \\
Fe & -4.57(13;79) & -4.57(12;96) & -4.59  \\
Ni & -5.84(21;16) & -5.78(22;16) & -5.81  \\
Zn & -8.04(23;2)  & -8.06(29;2)  & -7.44  \\
Y  & -9.45(21;3)  & -9.53(19;3)  & -9.83  \\
Ba & -8.95(20;4)  & -9.00(06;3)  & -9.87  \\
\tableline
\end{tabular}
\end{center}
\label{tab-hda-abundances}
\end{table}
\begin{table}
\caption{As in Table~\ref{tab-hda-abundances}, but for \bda. The Li abundance 
is an upper limit.}
\begin{center}
\begin{tabular}{lccc}
\tableline\tableline
        & \multicolumn{3}{c}{$\log (N/N_{\rm tot})$} \\ 
Element & SOPHIE & HARPS & Sun \\
\tableline
Li & -(-;0)        & $<$-9.80(-;1)    & -10.99 \\
C  & -3.58(12;5)   &    -3.60(15;3)   & -3.65 \\
Na & -5.29(-;1)    & 	-5.43(04;2)   & -5.87 \\
Mg & -4.69(17;6)   & 	-4.62(18;6)   & -4.51 \\
Si & -4.15(13;7)   & 	-4.37(19;9)   & -4.53 \\
S  & -4.60(-;1)    & 	-4.60(09;4)   & -4.90 \\
Ca & -6.19(21;15)  &    -6.18(10;19)  & -5.73 \\
Sc & -10.05(30;7)  &   -10.18(38;4)   & -8.99 \\
Ti & -7.20(28;18)  &    -7.28(15;20)  & -7.14 \\
Cr & -6.33(25;20)  &    -6.40(18;21)  & -6.40 \\
Mn & -6.69(18;8)   & 	-6.67(21;7)   & -6.65\\
Fe & -4.52(18;204) &    -4.53(15;110) & -4.59\\
Ni & -5.53(18;25)  &    -5.50(10;19)  & -5.81\\
Zn & -7.32(10;2)   & 	-7.43(01;2)   & -7.44\\
Y  & -9.56(18;3)   & 	-9.56(07;5)   & -9.83\\
Ba & -9.57(08;2)   & 	-9.61(02;3)   & -9.87\\
\tableline
\end{tabular}
\end{center}
\label{tab-bda-abundances}
\end{table}

The uncertainty of \Tf\ was derived directly from hydrogen line fitting with 
\errTeff\ of 200\,K for both stars. Due to the rather low \Tf\ 
and high \vsini\ of the two stars, only a few \ion{Fe}{2} lines could be 
measured compared to \ion{Fe}{1} lines resulting in a \errlogg\ of 0.2. A smaller 
error for \lgg\ was derived for \hda\ using its parallax and evolutionary models as was 
described at the beginning of this section. Iron was the only element for which we were 
able to use enough lines of two ionization stages for determining \lgg. 

\begin{figure}
    \epsscale{0.70}
    \plotone{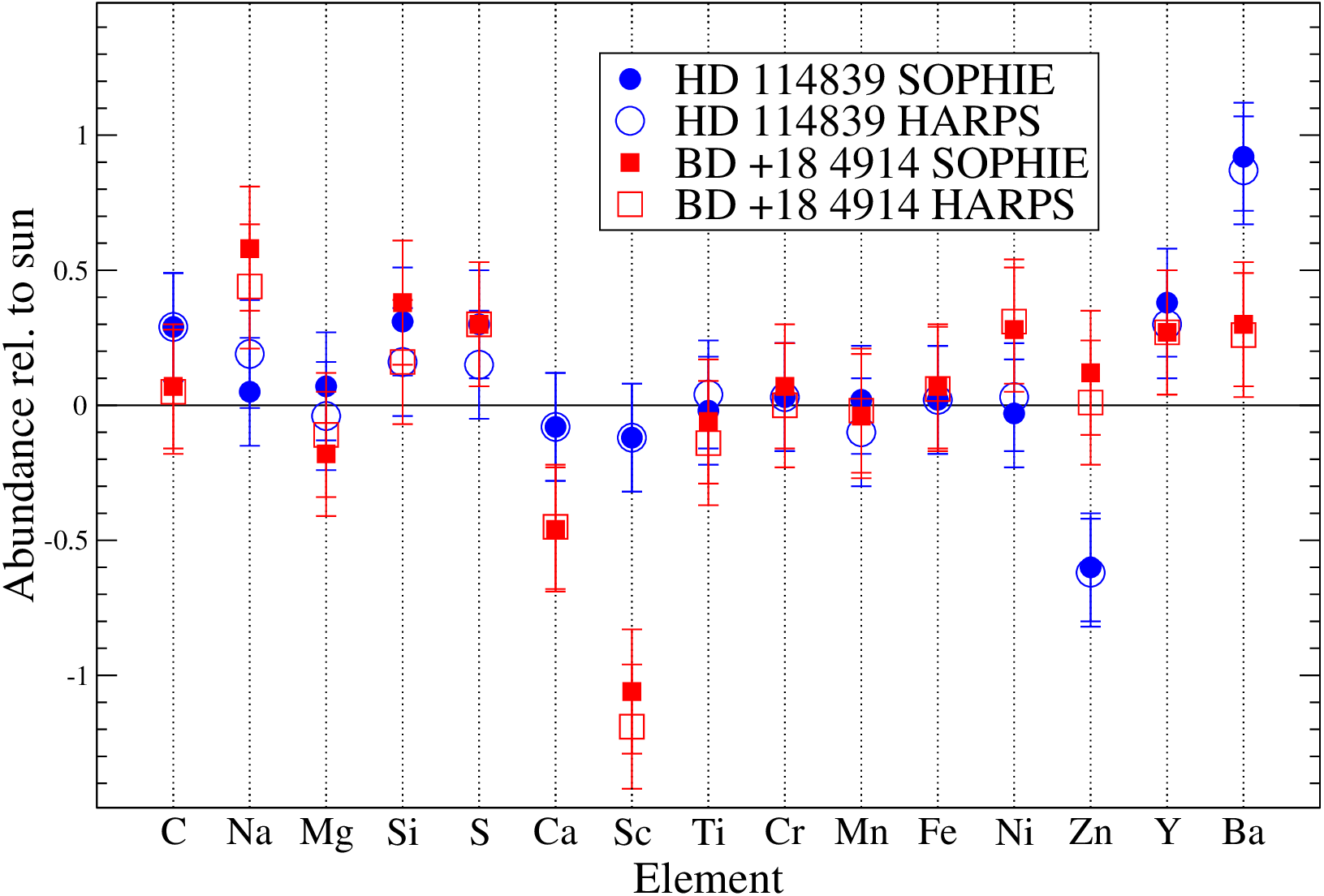}
 \caption{Comparison between the abundances relative to the Sun 
 \citep{asplundetal2005} obtained for \hda\ (circles) and \bda\ (squares) 
 and derived on the basis of the SOPHIE (full symbols) and HARPS (open 
 symbols) spectra. The abundance error bars are explained in the text and are 
 0.20\,dex for \hda\ and 0.23\,dex for \bda, respectively.}
 \label{fig-abundances}
\end{figure}

The abundance uncertainties listed in Tables~\ref{tab-hda-abundances} and 
\ref{tab-bda-abundances} are the standard deviations of the mean abundances 
derived from the individual line abundances, which include also the errors 
from oscillator strengths and continuum normalization \citep{fossati09}.
The standard deviations of the mean abundances underestimate the actual 
abundance uncertainty, since uncertainties of the fundamental parameters 
should also be taken into account. 
The error bars used for Fig.\,\ref{fig-abundances} were computed from the 
square-root of the sum of squared abundance changes resulting from varying 
fundamental parameters by 1$\sigma$.

For iron, e. g., the abundance changes by 0.12\,dex, if we change \Tf\ by 
1$\sigma$ (= 200\,K). In the case of \vmic\ the abundance changes by 0.09\,dex 
when changing \vmic\ by 1$\sigma$ (= 0.7\,\kms), while the uncertainty of \lgg\ 
increases the abundance error only by 0.01\,dex. All these uncertainties add up 
for the iron abundance to 0.20\,dex for \hda, and to 0.23\,dex for \bda.

The results we obtained from the SOPHIE and HARPS spectra are in comfortably good
agreement, confirming our continuum normalization, determination of the fundamental parameters and 
abundances, and their uncertainties, as shown in Fig.~\ref{fig-abundances}. Note that the same lines have not always been selected
for the same star observed with the two spectrographs.

\begin{table}
\caption{Final parameters of the program stars.}
\begin{center}
\begin{tabular}{lcccc}
\tableline\tableline
  & \Tf\ & \lgg\ & \vmic & \vsini\ \\
  & [K]  & [\cms] & [\kms] & [\kms]  \\
\tableline
\hda\ &  7100$\pm$200 & 4.0$\pm$0.15 & 2.7$\pm$0.5  & 68$\pm$2 \\
\bda\ &  6900$\pm$200 & 3.8$\pm$0.20 & 3.2$\pm$0.7  & 38$\pm$2 \\
\tableline
\end{tabular}
\end{center}
\label{tab-finalparam}
\end{table} 

\section{Spectroscopic characteristics of \hda\ and \bda}\label{Am nature}

\begin{figure}
    \epsscale{0.70}
    \plotone{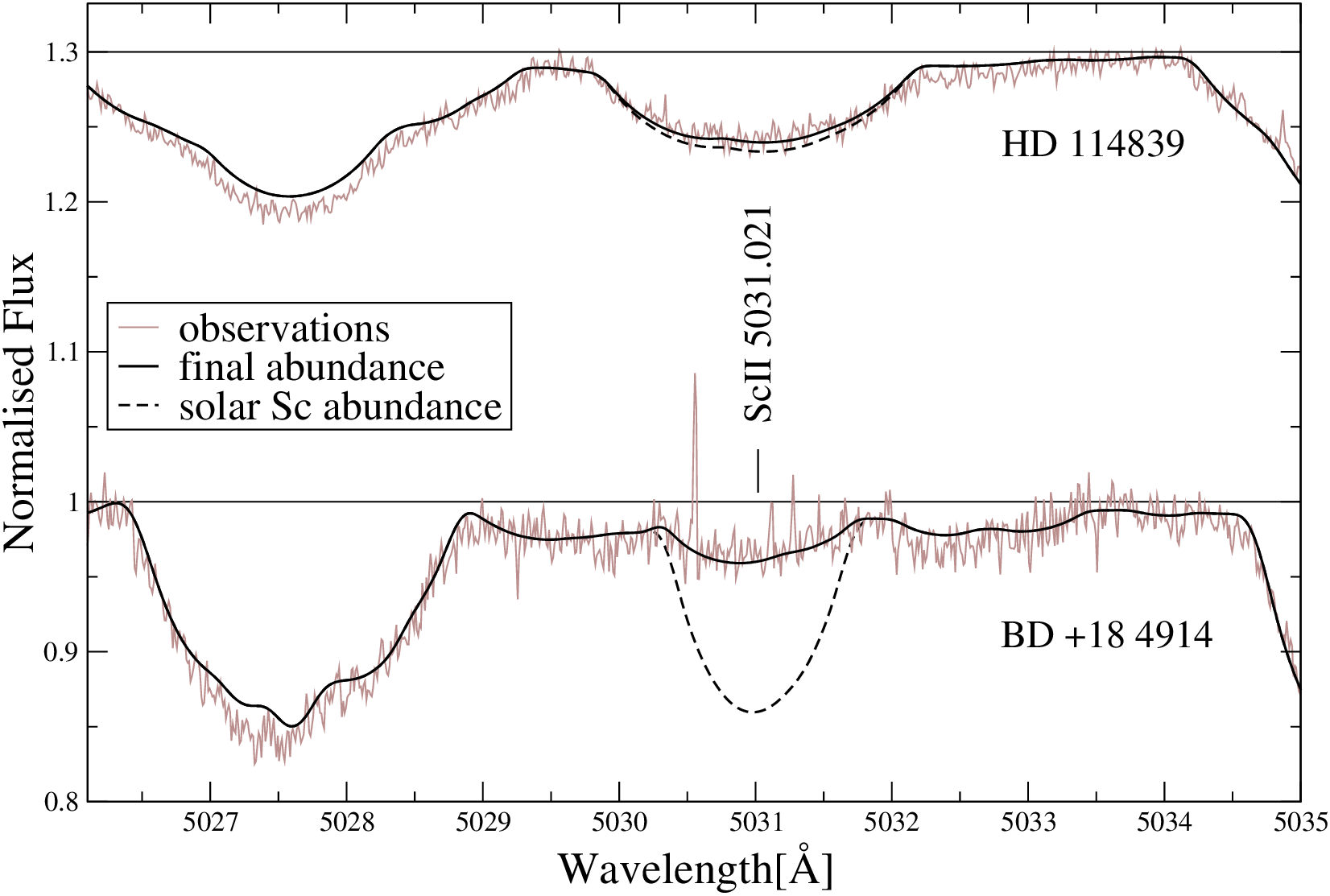}
 \caption{Observed SOPHIE spectrum of \hda\ and HARPS spectrum of \bda\ compared to synthetic spectra around 
the \ion{Sc}{2} line at 5031\,\AA\ and of three iron lines at 5027.5\,\AA. The dashed line is for solar Sc abundance 
and the full lines are for the abundance derived by us. The spectra for \hda\ are offset by 0.3 for better visibility.}
 \label{fig-hda-scline}
\end{figure}

Figure~\ref{fig-hda-scline} shows the spectral region
around the \ion{Sc}{2} line at $\lambda$5031\,\AA\ and of three iron lines at 5027.5\,\AA\ for the two stars in comparison with synthetic 
spectra calculated once with the Sc solar abundance (dashed line) 
and a second time with the Sc abundance obtained individually for each star (full line). The star \bda\ shows clearly 
a strong Sc underabundance, while \hda\ presents a solar Sc abundance. This 
confirms \bda\ being an Am star, which is corroborated by the Ca
underabundance and the significant overabundance of Ni and Y.

\begin{figure}
    \epsscale{0.70}
    \plotone{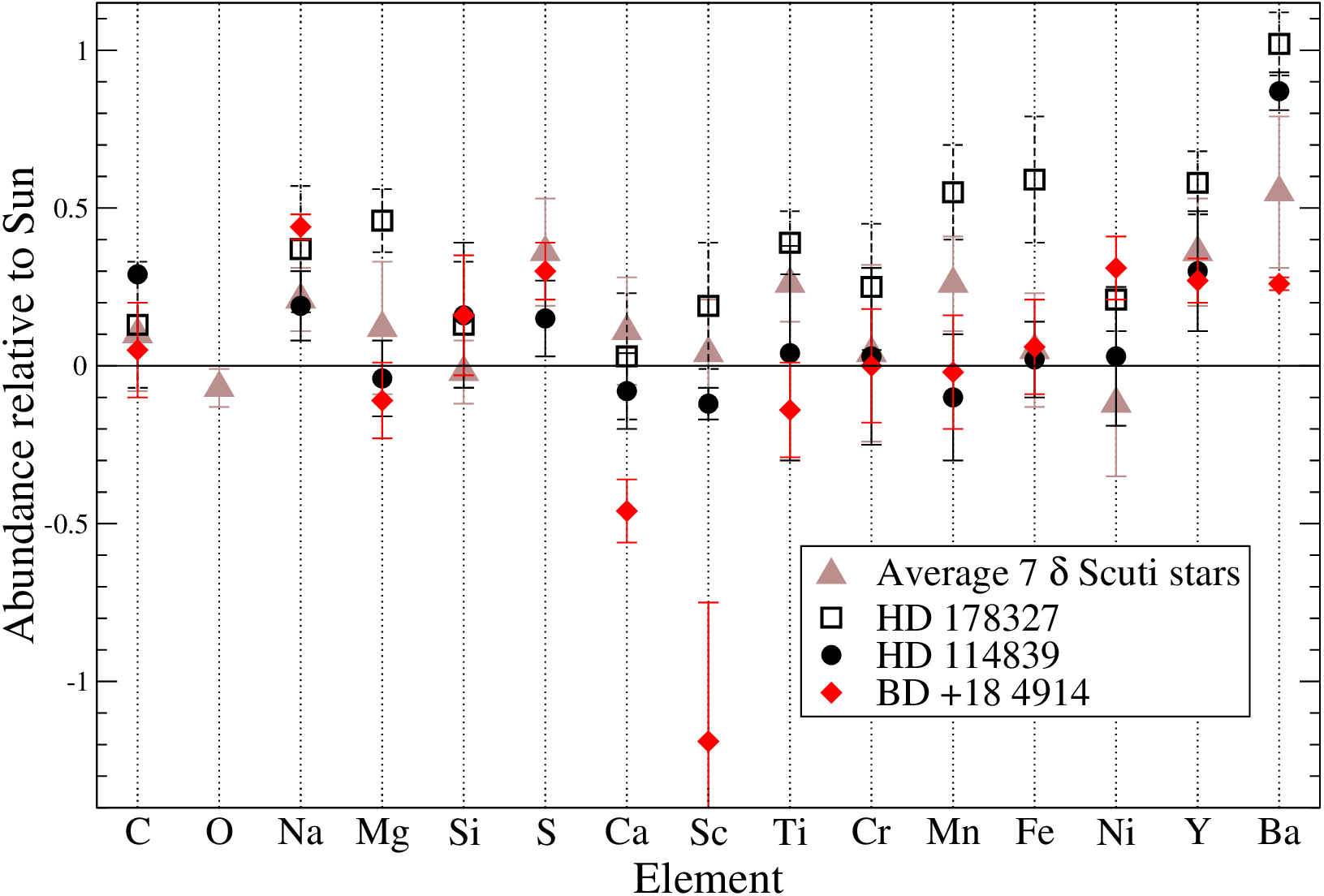}
 \caption{Mean abundances relative to the Sun of \hda\ (filled circles) and of
 \bda\ (filled rhombs), in comparison to the mean abundances of the sample of
 field \ds\ stars (triangles) published by \citet{Fossati08b} to HD\,178327 (open squares) by \citet{Balona11}. 
The error bars  correspond to the standard deviation of the mean abundance.}
 \label{fig-abundances-vs7ds}
\end{figure}

\begin{figure}
    \epsscale{0.70}
    \plotone{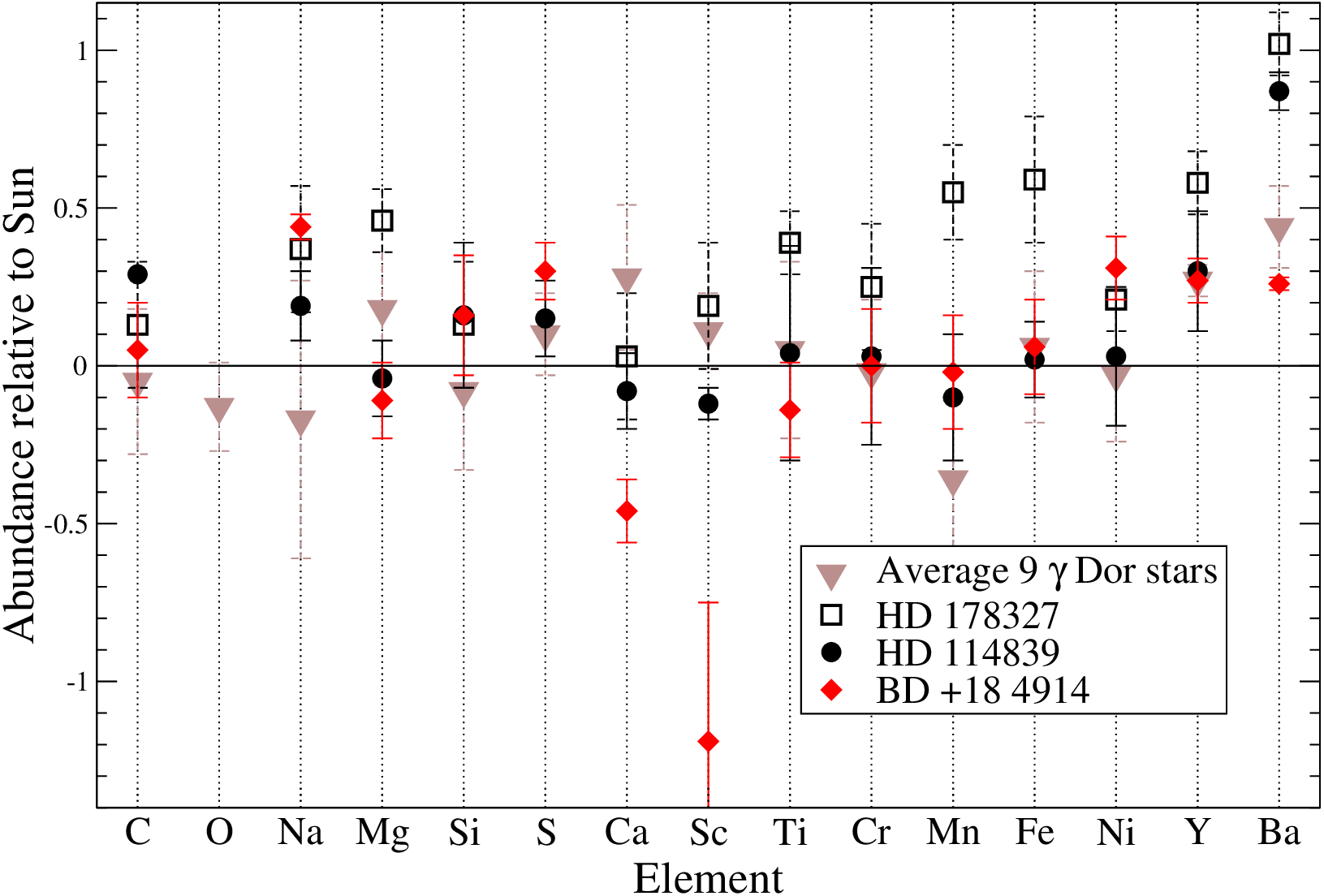}
 \caption{Same as Fig.\,\ref{fig-abundances-vs7ds} but the results are compared to a sample 
  of \gdor\ stars (inverted triangles) published by \citet{Bruntt08}. }
 \label{fig-abundances-vs9gd}
\end{figure}

In Fig.~\ref{fig-abundances-vs7ds} and \ref{fig-abundances-vs9gd} we compare the abundances obtained for 
\hda\ and \bda\ with the average abundances published by \citet{Fossati08b} 
for seven nearby field \ds\ stars. This comparison is straight forward 
because parameters and abundances were obtained with similar methods 
and codes. We do not find any systematic difference between the mean 
abundances of the field \ds\ stars and of \hda, while \bda\ clearly shows 
its Am nature. 
Additionally, the results were compared to that of 9 \gdor\ stars obtained by \citet{Bruntt08}. 
These authors did not find any typical abundance pattern among \gdor\ stars, however they found a larger 
scatter in the abundances compared to the reference stars. With a reanalysis of two references stars used 
by \citet{Bruntt08}, \citet{fossati11} showed that the methodology for parameter determination adopted 
by \citet{Bruntt08} led them to erroneous results. Consequently, the abundances of the nine \gdor\  
stars might not be reliable and conclusions from this comparison should be taken with caution.

Very recently, \citet{Balona11} presented the analysis of 10 Am stars observed by Kepler of which HD\,178327 shows hybrid pulsation. 
But with a mild overabundance of Ca and Sc, HD\,178327 misses an important classification criterion for Am stars where the mentioned 
elements should be significantly underabundant. The abundances of 14 elements determined by \citet{Balona11} for this hybrid star 
are included in our Fig.\,\ref{fig-abundances-vs7ds} and \ref{fig-abundances-vs9gd}, hence increasing the total number of hitherto 
spectroscopically investigated hybrid stars to five.

\begin{figure}
    \epsscale{0.70}
    \plotone{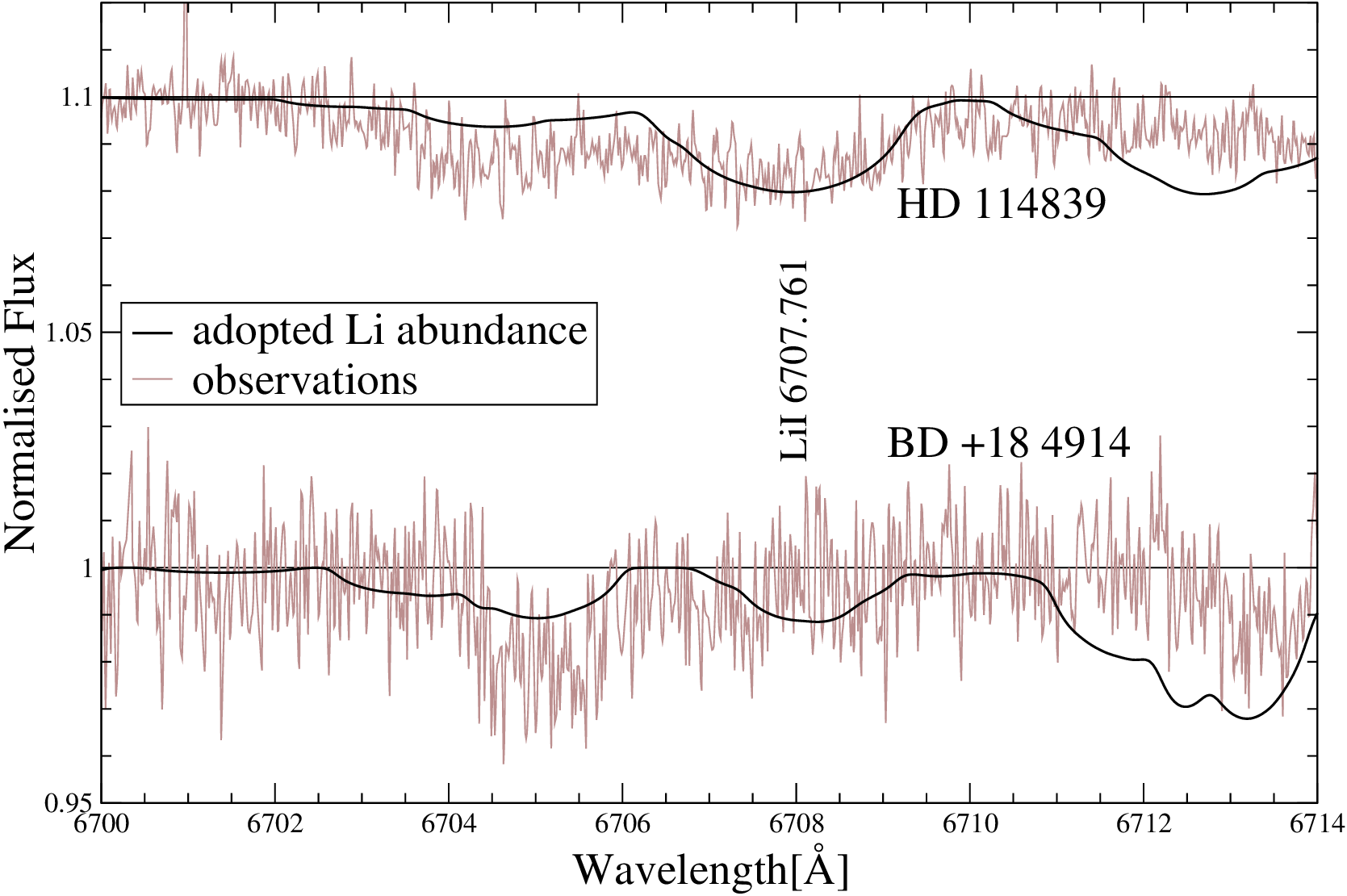}
 \caption{Top: observed HARPS spectrum of \hda\ compared to the synthetic spectrum (full black line) computed with the Li abundance derived by us in the 
region of the \ion{Li}{1} line at $\lambda$6707\,\AA.  Bottom: same as top, but for \bda\ and with the upper limit for the Li abundance. For 
better visibility the spectrum of \hda\ is offset by +0.1.}

 \label{fig-hda-bda-Li}
\end{figure}

Figure~\ref{fig-hda-bda-Li} shows the HARPS spectra in the region of the Li 
line at $\lambda$6707\,\AA\ in comparison with a synthetic 
spectrum calculated with the Li abundance obtained for \hda\ (top).  
For this star the Li line is clearly visible and leads to an abundance of -8.91\,dex.   
For \bda\ (bottom) the Li line is not visible and we obtain an upper limit of 
-9.80\,dex. The synthetic spectrum for \bda\ is computed with this abundance.

\citet{Burkhart05} and \citet{Fossati07} derived the Li abundance 
in a set of chemically normal A-type stars and compared it to Am 
stars, concluding that Am stars have a lower Li abundance compared to 
chemically normal A-type stars. The Li abundance we obtained for \hda\ and 
\bda\ reproduces this pattern and strengthens our classification of \hda\ as a
chemically normal star and of \bda\ as an Am star.

A very recent study of \ds, \gdor, and hybrid stars \citep{Uytterhoeven11} observed 
by the space mission {\it Kepler} indicates that possibly two more hybrid candidate stars are chemically peculiar: HD\,181206 (Ap or Am) and HD\,178327 (Am). 
For 58 other hybrid candidate stars spectral types 
based on published spectral classifications are available, but with no indication of peculiarity.

\section{Evolutionary status of \hda\ and \bda}\label{hrd}

\begin{figure}
    \epsscale{0.70}
    \plotone{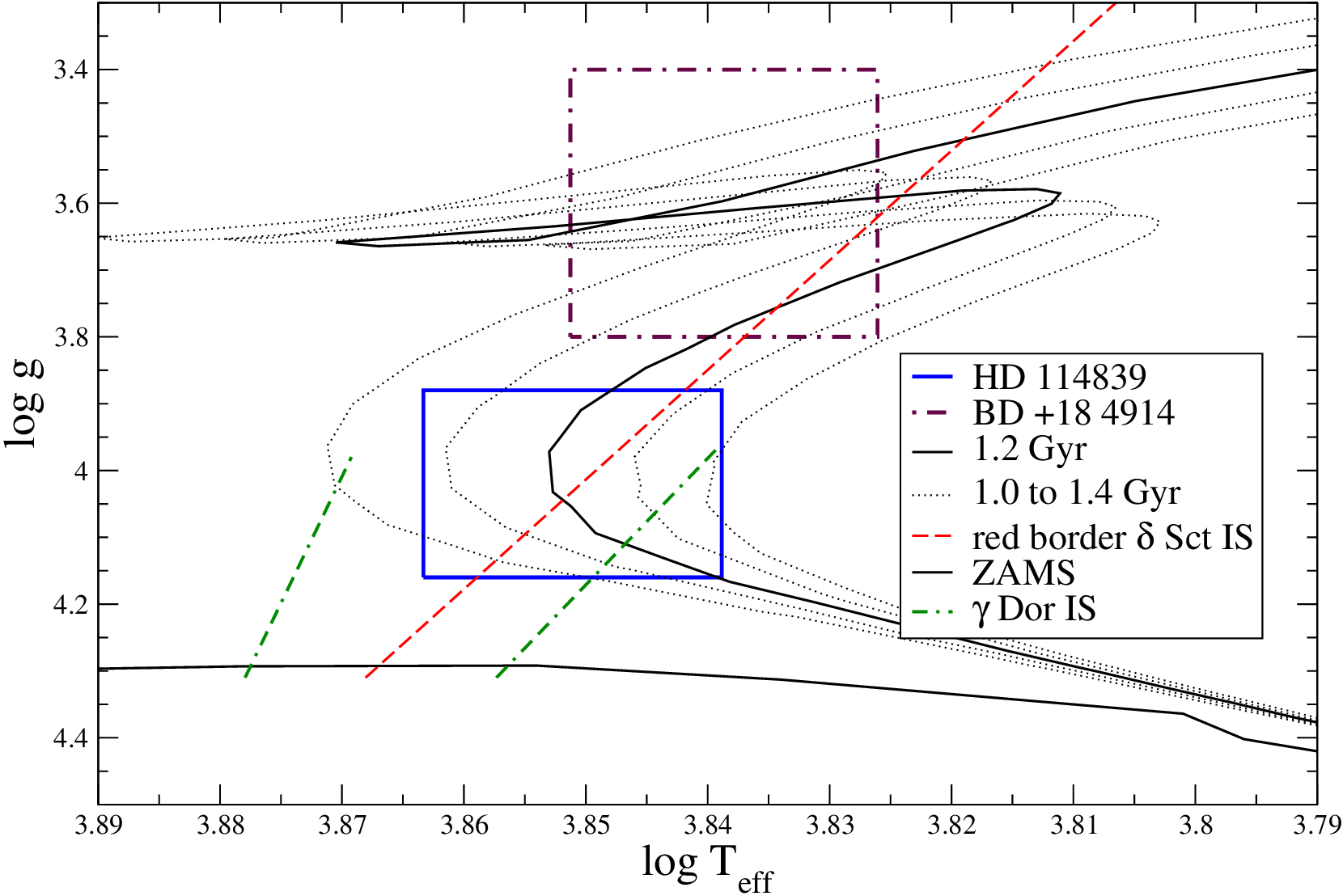}
 \caption{Position of \hda\ (full line errorbox) and \bda\ (dash-dotted line
 errorbox) in the \Tf-\lgg\ diagram, in comparison with solar metallicity
 isochrones by \citet{marigo}. The isochrone crossing the center of the
 \hda\ errorbox (full line) corresponds to a 1.2\,Gyr old isochrone. The dotted
 line represent 1.0, 1.1, 1.3, and 1.4\,Gyr isochrones, from left to right.
 The Zero Age Main Sequence is also indicated by a full line. The dashed line
  indicates the red edge of the \ds\ instability strip, while the dash-dotted lines show the 
 blue and red edge of the \gdor\ instability strip. Both stars fall inside 
 the \ds-\gdor\ instability strip.}

 \label{fig-bda-hda-isochrones}
\end{figure}

Fig.\,\ref{fig-bda-hda-isochrones} shows \hda\ and \bda\ in the \Tf-\lgg\ plane with the 
respective error boxes derived in Sec.\,\ref{s:param}. The figure also shows isochrones 
from \citet{marigo}, for ages between 1 and 1.4 Gyr, from left to right. Solar 
metallicity was chosen, because we obtained solar Fe abundance for both stars. The error boxes 
for both stars cover part of the cool \ds\ and the \gdor\ instability regions and a range of 
evolutionary stages, masses and radii (and thereby mean densities). 

We used a grid of A-F models\footnote{This model grid is part of the Spanish Virtual 
Observatory Project: VOTA (Virtual Observatory Tools for asteroseismology, Su\'arez et al. in prep.) 
to  be released during 2011.} provided by one of us (JCS) using the evolutionary code \emph{CESAM} \citep{Morel97} to explore 
the given parameter space in more detail. We used different values for the mass, convection efficiency 
and overshoot parameters within ranges typical for these stars, \citep[see e.g. ][for \ds\ and \gdor\ 
stars]{Sua05a, Bruntt07, Rod06a, Rod06b}. Metallicity was fixed to the solar value. Numerics were 
optimized following the prescriptions suggested by \citet{Moya08} within ESTA/CoRoT activities\footnote{http://www.astro.up.pt/corot/}. 
Rotation may play an important role for detailed asteroseismic studies of these stars \citep{Goupil05, Sua06a}, 
but was not taken into account for the used models. 

We found representative models for \hda\ with masses between 1.52 and 1.71 \Msun, radii between 1.72 and 
2.26 \Rsun, and ages between 905 and 1405 Myr, respectively. Those for \bda\ were found to have masses 
between 1.6 and 2.10 \Msun, radii between 2.69 and 4.43 \Rsun, and ages between 935 and 1782 Myr. In 
all cases, the best model fit was obtained for the mixing-length parameter ($\alpha_{\rm MLT})=0.5$, which is the value found by 
\citet{Casas06} for this type of stars using asteroseismic techniques. Overshoot values range from 0.1 to 
0.3 for fitting models. 

\citet{Turcotte00} investigated the effect of diffusion on pulsation for 
upper main sequence stars. They concluded that young Am stars are stable 
against pulsation, but become unstable during evolution towards the red 
edge of the \ds\ instability strip. The chemical and evolutionary characteristics 
of \bda\ presented here are consistent with \citet{Turcotte00}'s conclusion.

\section{Conclusions}

Only one hybrid star, HD\,49434, was investigated spectroscopically in detail before we started our study of \gdhybrid s. 
However its hybrid nature is still uncertain and needs confirmation. HD\,49434 did not show chemical peculiarities. 
Hybrid pulsators and stable stars populate the same region of the HR diagram, a situation that is similar to 
roAp stars \citep{Ryabchikova04}. For the latter a clear spectroscopic signature could be detected which allows even 
prediction of pulsation in CP stars \citep{Kochukhov02}. A similar feature for hybrid pulsators would be extremely important. 

A link between the Am phenomenon and hybrid pulsation was speculated, because of the first detected hybrid, HD\,8801, being an Am star. 
Furthermore, early in the literature significant overabundances of iron and/or iron-peak elements were expected as spectroscopic 
indicators for the driving mechanism. However, our comparison with seven \ds\ \citep{Fossati08b} and nine \gdor\ stars \citep{Bruntt08} 
do not indicate any significant chemical difference between these groups of stars. 

With the present paper we nearly doubled the number of spectroscopically investigated hybrid stars, but have not yet succeeded 
in identifying a particular spectroscopic signature for this group of pulsating stars. For the evolved \bda\ we found 
an abundance spectrum typical for an Am star, but could not confirm this peculiarity for the less evolved star \hda, which 
is classified in the literature a mild (or marginal) Am star.

\gdhybrid s pulsate simultaneously in p- and g-modes, which are sensitive to the envelope as well as to the core. Hence, they 
are key objects for investigating A and F type stars with asteroseismic techniques. Accurate values for temperature, gravity and 
chemical composition are needed as boundary conditions for stellar models used for fitting the observed eigenfrequencies. We provide 
such boundary conditions for a seismic modeling of \hda\ and \bda, as well as \vsini\ and \vmic.

A statistically significant spectroscopic investigation of \gdhybrid\ stars still is missing, but would be rewarding considering 
the asteroseismic potential of this group of pulsators.


\acknowledgments

The authors gratefully acknowledge the development of the tools by C. St\"utz and D. Shulyak, 
used for this research and to T. L\"uftinger for valuable discussion.
This project was supported by the Austrian Fonds zur F\"orderung der 
wissenschaftlichen Forschung (FWF, project 
P 22691-N16) and by an STFC Rolling Grant (LF). 
EP acknowledges support from the PRIN-INAF 2010 {\it Asteroseismology: looking 
inside the stars with space and ground-based observations}.
JCS was supported by the "Instituto de Astrof\'{\i}sica de
Andaluc\'{\i}a (CSIC)" by an "Excellence Project post-doctoral fellowship" financed 
by the Spanish "Conjerer\'{\i}a de Innovaci\'on, Ciencia y Empresa de la Junta de Andaluc\'{\i}a" under
project "FQM4156-2008".
KU acknowledges financial support by the Deutsche  
Forschungsgemeinschaft (DFG) in the framework of project UY 52/1-1,  
and by the Spanish National Plan of R\&D for 2010, project  
AYA2010-17803.
This work made use of the SIMBAD data base, operated at the CDS, Strasbourg, France, and is based on observations 
collected at the ESO 3.6\,m telescope at La Silla (Chile), and at the Observatoire de Haute Provence (France). 
We acknowledge the OPTICON programme (Ref number: 2008/053). 

{\it Facilities:} \facility{OHP:1.93m (SOPHIE)} \facility{ESO:3.6m (HARPS)}.


\begin{thebibliography}{}
\bibitem[Asplund et al.(2005)]{asplundetal2005}
	Asplund, M., Grevesse, N., \& Sauval, A.~J., 2005,
	Astronomical Society of the Pacific Conference Series, 336, 25
\bibitem[Auri{\`e}re et al.(2010)] {auriere10}
        Auri{\`e}re et al., 2010, \aap\, 523, 40
\bibitem[Baglin et al.(2006)] {esa3} 
	Baglin, A., Auvergne, M., Barge, P., et al., 2006, in ``The CoRoT Mission, 
	Pre-Launch Status, Stellar Seismology and Planet Finding'', ed. M.~Fridlund, 
	A.~Baglin, J.~Lochard, \& L.~Conroy, ESA Publications Division, Nordwijk, 
	Netherlands, ESA SP-1036, 33
\bibitem[Balona (1994)]{Balona94}
	Balona, L.~A., 1994, \mnras, 268, 119
\bibitem[Balona et al.(2011)]{Balona11}
	Balona, L.~A. et al., 2011, \mnras, 414, 792
\bibitem[Barklem et al.(2002)]{barklem02}
	Barklem, P.~S., Stempels, H.~C., Allende Prieto, C., Kochukhov, O. P., Piskunov, N. \& O'Mara, B. J, 2002, \aap,
	385, 951
\bibitem[Borucki et al.(2010)] {Borucki2010}
        Borucki, W. J. et al., 2010, Science, 327, 977
\bibitem[Breger (2000)]{Breger00}
	Breger, M. 2000, ASPC, 210, 3
\bibitem[Bruntt et al.(2007)]{Bruntt07}
	Bruntt, H. et al., 2007, \aap, 461, 619
\bibitem[Bruntt et al.(2008)] {Bruntt08}
        Bruntt, H., De Cat, P. \& Aerts, C., 2008, \aap, 478, 487
\bibitem[Burkhart et al.(2005)]{Burkhart05}
	Burkhart, C., Coupry, M.~F., Faraggiana, R., \& Gerbaldi, M., 
        2005, \aap, 429, 1043
\bibitem[Casas et al.(2006)]{Casas06}
	Casas, R., Su\'arez, J. C., Moya, A. \& Garrido, R., 2006, \aap, 455, 1019
\bibitem[Charbonneau \& Michaud (1991)]{charbonneau}
	Charbonneau, P. \& Michaud, G., 1991, \apj, 370, 693
\bibitem[Canuto \& Mazzitelli (1991a)]{cm1}
        Canuto, V. M., \& Mazzitelli, I., 1991, \apj, 370, 295
\bibitem[Canuto \& Mazzitelli (1991b)]{cm2}
        Canuto, V. M., \& Mazzitelli, I., 1992, \apj, 389, 724
\bibitem[Clausen \& Jensen (1979)]{Clausen79}
	Clausen, J.~V. \& Jensen, K.~S., 1979, IAU Colloq. 47, 9, 479
\bibitem[Flower (1996)] {Flower96}
        Flower, P. J., 1996, ApJ, 496, 355
\bibitem[Fossati et al.(2007)]{Fossati07}
	Fossati, L., Bagnulo, S., Monier, R., Khan, S. A., Kochukhov, O., Landstreet, J., Wade, G. \& Weiss, W.,
	2007, \aap, 476, 911 
\bibitem[Fossati et al.(2008a)]{Fossati08} 
	Fossati, L., Bagnulo, S., Landstreet, J., Wade, G., Kochukhov, O., Monier, R., Weiss, W. \& Gebran, M.
	2008b, \aap, 483, 891
\bibitem[Fossati et al.(2008b)]{Fossati08b}
	Fossati, L., Kolenberg, K., Reegen, P. \& Weiss, W. 2008a, \aap, 485,
	257
\bibitem[Fossati et al.(2009)]{fossati09}
	Fossati, L., Ryabchikova, T., Bagnulo, S., Alecian, E., Grunhut, J., Kochukhov, O. \& Wade, G, 2009, \aap, 503, 945
\bibitem[Fossati et al.(2011)]{fossati11}
        Fossati, L., Ryabchikova, T., Shulyak, D.~V., Haswell, C.~A., Elmasli, A., Pandey, C.~P., Barnes, T.~G. \& Zwintz, K. 2011, MNRAS, in press (arXiv: 1106.4406) 
\bibitem[Grigahc{\`e}ne et al.(2010)]{Grig10} 
  Grigahc{\`e}ne, A. et al., 2010, \apj, 713, 192
\bibitem[Goupil et al.(2005)]{Goupil05}
	Goupil, M.-J., Dupret, M. A., Samadi, R., Boehm, T., Alecian, E., Suarez, J. C., Lebreton, Y. \& Catala, C., 2005, JA\&A, 26, 249 
\bibitem[Hauck \& Mermilliod(1998)]{HauckM}
	Hauck, B., \& Mermilliod, M., 1998, \aaps, 129, 431
\bibitem[Hareter et al.(2010)]{Hareter10} 
	Hareter, M., et al.2010, AN, P49  (arXiv:1007.3176)
\bibitem[Heiter et al.(2002)]{heiter}
        Heiter, U., et al.2002, \aap, 392, 619
\bibitem[Henry \& Fekel (2005)]{Henryfekel05}
	Henry, G.~W. \& Fekel, F.~C., 2005, \aj, 129, 2026
\bibitem[Hill et al.(1976)]{Hill76}
	Hill, G., Allison, A., Fisher, W.~A., Odgers, G. J., Pfannenschmidt, E. L., Younger, P. F. \& Hilditch, R. W., 1976, \memras, 82, 69
\bibitem[Iliev \& Budaj (2008)] {Iliev2008}
	Iliev, I. Kh. \& Budaj J., 2009, Contrib. Astron. Obs. Skalnate Pleso, 38, 2, 129
\bibitem[Kaye et al.(1999)]{Kaye99}
	Kaye, A.~B., Handler, G., Krisciunas, K., Poretti, E. \& Zerbi, F.~M., 1999, \pasp, 111, 840
\bibitem[King et al.(2006)]{King06}
	King, H., et al., 2006, CoAST, 148, 28
\bibitem[Kochukhov et al.(2002)]{Kochukhov02}
	Kochukhov, O., Landstreet, J.~D., Ryabchikova, T., Weiss, W. W. \& Kupka, F., 2004, \mnras, 337, 1
\bibitem[Kochukhov (2007)]{Kochukhov07}
	Kochukhov, O. 2007,  
	Spectrum synthesis for magnetic, chemically stratified stellar 
	atmospheres, Physics of Magnetic Stars, 109, 118
\bibitem[Kupka et al.(1999)]{vald2}
        Kupka, F., Piskunov, N., Ryabchikova, T. A., Stempels, H. C., \& 
	Weiss, W. W. 1999, \aaps, 138, 119
\bibitem[Kurucz (1993a)]{Kurucz93b} 
	Kurucz, R. 1993, ATLAS9: Stellar Atmosphere Programs and 2 km/s 
	grid.~Kurucz CD-ROM No.~13 (Cambridge: Smithsonian Astrophysical 
	Observatory)
\bibitem[Marigo et al.(2008)]{marigo}
	Marigo, P., Girardi, L., Bressan, A., Groenewegen, M. A. T., Silva, L. \& Granato, G. L., 2008, \aap, 482, 883
\bibitem[Mathias et al.(2004)]{Mathias04}
        Mathias et al., 2004, \aap, 417, 189	
\bibitem[Michaud (1970)] {Michaud70}
	Michaud, G., 1970, ApJ, 160, 641
\bibitem[Morel (1997)] {Morel97}
	Morel, P., 1997, \aaps, 124, 597
\bibitem[Moya et al.(2008)]{Moya08}
	Moya, A., et al., 2008, \apss, 316, 231 
\bibitem[Piskunov et al.(1995)]{vald1}
        Piskunov, N. E., Kupka, F., Ryabchikova, T. A., Weiss, W. W., \& 
	Jeffery, C. S. 1995, \aaps, 112, 525
\bibitem[Pribulla et al.(2009)]{Pribulla09}
	Pribulla, T., Rucinski, S.~M., Kuschnig, R., Og{\l}oza, W. \& 
	Pilecki, B. 2009, \mnras, 392, 847
\bibitem[Rodr\'iguez et al.(2006a)]{Rod06a}
	Rodr\'iguez, E., et al., 2006, \aap, 456, 261
\bibitem[Rodr\'iguez et al.(2006b)]{Rod06b}
	Rodr\'iguez, E., et al., 2006, \aap, 450, 715
\bibitem[Rowe et al.(2006)]{Rowe06} 
	Rowe, J.~F., et al., 2006, CoAST, 148, 34
\bibitem[Ryabchikova et al.(1999)]{vald3} 
	Ryabchikova, T.~A., Piskunov, N.~E., Stempels, H.~C., Kupka, F., 
	\& Weiss, W.~W. 1999, Phis. Scr., T83, 162
\bibitem[Ryabchikova et al.(2004)]{Ryabchikova04}
        Ryabchikova, T., Nesvacil, N., Weiss, W. W., Kochukhov, O. \& St\"utz, Ch., 2004, \aap, 423, 705
\bibitem[Su\'arez et al.(2005a)]{Sua05a}
	Su\'arez, J. C., Bruntt, H. \& Buzasi, D., 2005, \aap, 438, 633
\bibitem[Su\'arez et al.(2006)]{Sua06a}
	Su\'arez, J. C., Goupil, M. J., \& Morel, P.,  2006, \aap, 449, 673
\bibitem[Turcotte et al.(2000)]{Turcotte00} 
	Turcotte, S., Richer, J., Michaud, G. \& Christensen-Dalsgaard, J. 2000,
	\aap, 360, 603
\bibitem[Uytterhoeven et al.(2008)]{Uytterhoeven08}
	Uytterhoeven, K., et al., 2008, \aap, 489, 2213
\bibitem[Uytterhoeven et al.(2011), in press]{Uytterhoeven11}
	Uytterhoeven, K.,et al., 2011, \aap, in press (arXiv:1107.0335)
\bibitem[Van Leeuwen(2007)]{vanLeeuwen07}
	van Leeuwen, F. 2007, 
	Hipparcos, the New Reduction of the Raw Data. 
	Institute of Astronomy, Cambridge University, Cambridge, 
	UK Series: Astrophysics and Space Science Library, 
	Vol.~ 350 20 Springer Dordrecht
\bibitem[Walker et al.(2003)] {Walker03}
	Walker, G., et al., 2003, PASP, 115, 1023
\end{thebibliography}
\end{document}